\documentstyle[aps,prb,twocolumn,bezier,array,floats]{revtex}
\begin{document}
\title{
    About common physical nature of divergence and interference.
}
%
\author{ Daniel~L.~Miller }
\address{
  Intel Electronics Ltd.
  P.O.Box 3173, Jerusalem 91031, Israel\\
  e-mail:  Daniel.Miller@Intel.com
}

\date{ August 7, 2000 }
\maketitle
\begin{abstract}
   \widetext\leftskip=0.10753\textwidth \rightskip\leftskip
   The difference of actions between nearly lying unstable ``8'' and ``0''
   like trajectories in two dimensions is computed by methods of chaotic
   dynamics.  We found it to be equal to $pY\Theta/2$, where $\Theta$ is
   angle between paths at cross, $Y$ is distance between paths, and $p$ is
   momentum.  This kind of periodic trajectories contributes $\tau^2$
   (interference) term to spectral form-factor.
\end{abstract}

\narrowtext

\section{ Introduction }

Let's consider two particles inside infinite potential well, for example
two-dimensional billiard. What the matter do they interfere or not? In two
slit experiment one can measure intensity vs number of open slits and see
effect of phase difference between paths. The negative magneto-resistance of
disordered systems is manifestation of interference effects destroyed by
magnetic field\cite{AKLPL-80}.  The interference between particles inside
chaotic system suffers from lack of clear formulation, however it sounds
like
something that probably exist.

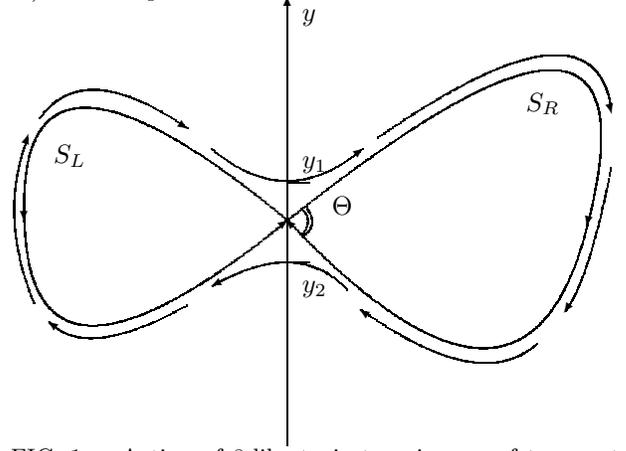
\begin{figure}
\vspace*{0.45in}
\unitlength 1.00mm
\linethickness{0.4pt}
\begin{picture}(90.00,70.00)
\put(40.00,0.00){\vector(0,1){60.00}}
\put(80.00,30.00){\vector(-1,-4){0.2}}
\bezier{420}(40.00,30.00)(90.00,70.00)(80.00,30.00)
\put(40.00,30.00){\vector(-1,1){0.2}}
\bezier{328}(80.00,30.00)(72.00,-4.00)(40.00,30.00)
\put(5.00,30.00){\vector(0,-1){0.2}}
\bezier{308}(40.00,30.00)(3.67,60.00)(5.00,30.00)
\put(40.00,30.00){\vector(4,3){0.2}}
\bezier{288}(5.00,30.00)(5.67,2.00)(40.00,30.00)
\put(49.96,39.48){\vector(4,3){0.2}}
\bezier{104}(29.92,39.48)(39.91,31.04)(49.96,39.48)
\put(30.04,21.30){\vector(-2,-1){0.2}}
\bezier{92}(47.92,20.80)(41.34,27.75)(30.04,21.30)
\bezier{20}(41.90,31.41)(43.51,30.11)(41.90,28.06)
\bezier{24}(42.39,31.85)(44.32,29.36)(42.02,27.88)
\put(46.00,32.00){\makebox(0,0)[lc]{$\Theta$}}
\put(42.00,37.00){\makebox(0,0)[lb]{$y_1$}}
\put(42.00,22.00){\makebox(0,0)[lt]{$y_2$}}
\put(9.00,40.00){\makebox(0,0)[lt]{$S_L$}}
\put(74.00,47.00){\makebox(0,0)[ct]{$S_R$}}
\put(42.00,57.00){\makebox(0,0)[lc]{$y$}}
\put(83.00,45.00){\vector(1,-4){0.2}}
\bezier{204}(52.00,41.00)(80.46,60.78)(83.00,45.00)
\put(77.00,18.00){\vector(-2,-3){0.2}}
\bezier{80}(83.00,37.00)(80.18,22.53)(77.00,18.00)
\put(49.80,18.11){\vector(-3,2){0.2}}
\bezier{120}(73.14,13.56)(65.54,6.79)(49.80,18.11)
\put(8.36,16.41){\vector(-4,3){0.2}}
\bezier{88}(26.57,18.67)(13.45,11.31)(8.36,16.41)
\put(5.43,40.94){\vector(1,3){0.2}}
\bezier{96}(6.28,18.96)(1.47,30.75)(5.43,40.94)
\put(26.66,42.26){\vector(3,-2){0.2}}
\bezier{108}(7.04,43.58)(12.89,51.79)(26.66,42.26)
\put(40.00,35.00){\line(1,0){3.00}}
\put(39.96,24.43){\line(1,0){3.02}}
\end{picture}
   \caption{ Action of 8-like trajectory is sum of two parts. There are
   two ways to connect left and right parts.
}
\label{fig:80geomtr}
\end{figure}

Physics of interference can be understood by simple methods of real space
trajectories. Postponing literature review, let's take one closed trajectory
crossing itself in real space under small angle $\Theta\ll1$. We prove that
this ``8'' like trajectory is accompanied by ``0'' like trajectory, see
construction below.  Their actions are simply related
\begin{equation}
    \Delta S \equiv S_0 - S_8 = { pY\Theta \over 2}
    = {p^2\over 2 }
    { m_{11} + m_{22} - 2m_{12}
        \over
    m_{11} m_{22} - m_{12}^2 } \Theta^2
\label{eq:new.1}
\end{equation}
where $m_{ij}$ are second derivatives of $S_0$ with respect to trajectory
displacements near cross, and $p$ is momentum. Figure~\ref{fig:80geomtr}
explains the choice of $\Theta$ and $Y=y_2-y_1$. Quantum particle on ``8''
like trajectory (we use language of semiclassical quantum mechanics ) will
interfere destructively with particle on ``0'' like trajectory. Indeed, i)
the phase difference between two wave functions is $\pi + \Delta S/\hbar $
because wave on ``8''-like trajectory must be inverted twice loosing $\pi/2$
phase each time and ii) the action difference $\Delta S$ can be
arbitrary small.

We arrive at subtle point. We do not mean that particles jump back and force
between ``8'' and ``0'' like trajectories. We just state that $\Delta S$ can
be arbitrary small. The closer trajectory approaches itself the smaller
$\Delta S$ is. The spectral form-factor, for example, is very sensitive to
trajectories with small phase difference. By making use of
Eq.~(\ref{eq:new.1}) and periodic orbit quantization
rule\cite{Gutzwiller-mar71} we can compute the so-called interference or
$\tau^2$ term of the spectral form-factor $K(\tau)$ (the diagonal term was
computed by Berry\cite{Berry-feb85}, see Eq.~(60) of the cited paper for
definition of the spectral form-factor).  First conclusion:  {\em The
interference in quantum chaos means that there are infinitely many pairs of
periodic trajectories with small difference of actions. }

\begin{figure}
\unitlength 1.00mm
\linethickness{0.4pt}
\begin{picture}(65.00,60.00)
\put(40.00,0.00){\vector(0,1){60.00}}
\put(5.00,30.00){\vector(0,-1){0.2}}
\bezier{308}(40.00,30.00)(3.67,60.00)(5.00,30.00)
\put(40.00,30.00){\vector(1,1){0.2}}
\bezier{288}(5.00,30.00)(5.67,2.00)(40.00,30.00)
\put(49.96,39.48){\vector(4,3){0.2}}
\bezier{104}(29.92,39.48)(39.91,31.04)(49.96,39.48)
\put(30.04,21.30){\vector(-2,-1){0.2}}
\bezier{92}(47.92,20.80)(41.34,27.75)(30.04,21.30)
\bezier{20}(41.90,31.41)(43.51,30.11)(41.90,28.06)
\bezier{24}(42.39,31.85)(44.32,29.36)(42.02,27.88)
\put(46.00,32.00){\makebox(0,0)[lc]{$\Theta$}}
\put(40.00,30.00){\vector(4,3){11.00}}
\put(48.00,22.00){\vector(-1,1){8.00}}
\put(53.00,41.00){\vector(4,3){12.00}}
\put(61.00,9.00){\vector(-1,1){10.00}}
\put(52.00,20.00){\vector(1,-1){11.00}}
\put(53.00,14.00){\makebox(0,0)[rt]{1}}
\put(57.00,17.00){\makebox(0,0)[lb]{reflected wave}}
\put(59.00,43.00){\makebox(0,0)[lt]{trasmitted wave}}
\end{picture}
   \caption{ Transmission and reflection of wave propagating
semiclassically
   along self-touching trajectory.
}
\label{fig:80trans}
\end{figure}
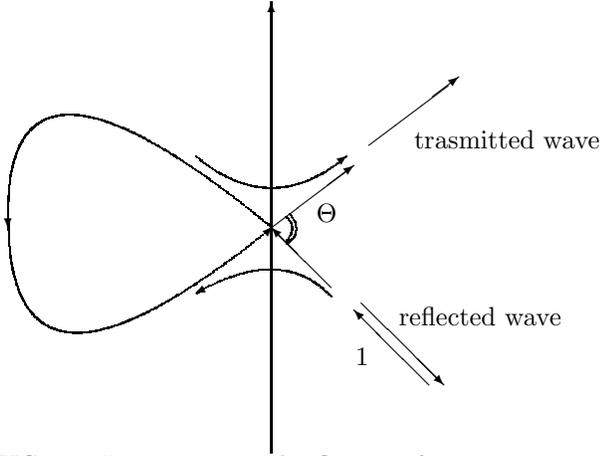

The mystery is that quantum waves do jump between nearly lying trajectories.
To understand how it works let's cut both ``8'' and ``0'' like trajectories
apart from the cross, see Fig.~\ref{fig:80trans}. It turns out that
transmission is $\sin^2(\Delta S/2\hbar)$ ( the trace of density propagator
mast be equal to absolute value square of trace of Green function). We
successfully arrive at second conclusion: {\em The interference in quantum
chaos means suppression of transmission of self-touching trajectories. }
This formulation of quantum interference is equivalent to that given by
Khmelnitskii for disordered systems.\cite{Khmelnitskii-84}

This introduction is followed by construction of pairs of near-lying
periodic
trajectories. As soon as we prove existence of such trajectories the
expression for the action difference Eq.~(\ref{eq:new.1}) can be simplified.
We will use the concept of self-correlated part of periodic
trajectory\cite{Aleiner-Larkin-96} and derive expression useful for the
operator form of the interference. The calculation of the $\tau^2$ term of
$K(\tau)$ is rather straightforward, but demand counting of 8-0 pairs of
trajectories of the certain period. It will be published elsewhere. We
review literature and conclude at the end.


\section{ Construction of near-lying trajectories }
\label{sec:con}

We broke the action of $8$-like trajectory to the two parts $ S_L + S_R$
and expand both on their end positions $y_1$ and $y_2$, see
Fig.~\ref{fig:80geomtr}:
\begin{eqnarray}
   S_L =
   S_L^{(0)}
   + {1\over 2} m^L_{11} y_1^2
   - m_{12}^{L} y_1 y_2
   + {1\over 2} m^L_{22} y_2^2
   + py_2\Theta
\label{eq:con.1}
\\
   S_R = S_R^{(0)}
   + {1\over 2} m^R_{11} y_1^2
   - m_{12}^{R} y_1 y_2
   + {1\over 2} m^R_{22} y_2^2
   - py_2\Theta
\label{eq:con.2}
\end{eqnarray}
The sum of actions has minimum $S_8 = S_L^{(0)} + S_R^{(0)} $ reached at
$y_1
= y_2 = 0$. If the particle follows left part of trajectory in other
direction, see Fig.~\ref{fig:80geomtr} then
\begin{equation}
   S_L' =
   S_L^{(0)}
   + {1\over 2} m^L_{11} y_2^2
   - m_{12}^{L} y_1 y_2
   + {1\over 2} m^L_{22} y_1^2
   + py_1\Theta \;.
\label{eq:con.3}
\end{equation}
The sum of actions $S_L' + S_R$ reaches it minimum at
\begin{equation}
   \vec Y = p\hat M^{-1} \vec \Theta
\label{eq:con.4}
\end{equation}
where $ \vec Y = (y_1; y_2) $, $ \vec \Theta = ( \Theta; -\Theta) $ and
\begin{equation}
   \hat M =
   \left( \begin{array}{cc}
   m_{11} & - m_{12} \\
   -m_{12} & m_{22}
   \end{array} \right)
   =
   \left( \begin{array}{cc}
   m^{L}_{22} + m^{R}_{11} & - m^{L}_{12} - m^{R}_{12} \\
   - m^{L}_{12} - m^{R}_{12} & m^{L}_{11} + m^{R}_{22}
   \end{array} \right) \;.
\label{eq:con.5}
\end{equation}
Here $m_{ij}$ without superscript are coefficients of expansion of 0-like
trajectory action $S_L' + S_R$ in terms of $y_1$ and $y_2$. The action of
the
true 0-like trajectory is
\begin{equation}
   S_0 = S_8 - \Delta S
   = S_8 - {p^2\over 2} \vec \Theta  \hat M^{-1} \vec \Theta \;,
\label{eq:con.6}
\end{equation}
and substitution of $m_{ij}$ gives Eq.~(\ref{eq:new.1}).

The present construction of nearly lying trajectory is valid under
assumptions that i) series Eqs.~(\ref{eq:con.1}) --  (\ref{eq:con.3})
exist  and ii) $Y$ is such small that third and higher derivatives
of actions are negligible. Clearly, the result Eq.~(\ref{eq:con.6}) does
not apply to all self-crossing trajectories, but to many of them. We leave
counting of such trajectories for separate publication.


\begin{figure}
\unitlength 1.00mm
\linethickness{0.4pt}
\begin{picture}(88.97,51.00)
\put(15.00,10.00){\line(2,1){61.00}}
\put(15.00,35.00){\line(4,-1){61.00}}
\bezier{368}(5.00,45.00)(51.00,26.00)(85.00,51.00)
\bezier{468}(5.00,0.00)(71.67,38.00)(84.00,0.00)
\put(12.00,42.00){\vector(1,2){0.2}}
\bezier{172}(12.00,4.00)(2.00,21.00)(12.00,42.00)
\put(16.92,34.53){\vector(1,2){0.2}}
\bezier{108}(16.19,10.58)(9.99,21.89)(16.92,34.53)
\put(73.99,39.41){\vector(-1,2){0.2}}
\bezier{88}(74.52,20.10)(79.25,30.09)(73.99,39.41)
\put(78.98,46.90){\vector(-1,2){0.2}}
\bezier{168}(78.59,10.51)(88.97,31.14)(78.98,46.90)
\put(76.00,48.00){\makebox(0,0)[rb]{$Y_0',\Theta_0'$}}
\put(75.00,33.00){\makebox(0,0)[rt]{$Y_8',\Theta_8'$}}
\put(17.00,28.00){\makebox(0,0)[lt]{$Y_8,\Theta_8$}}
\put(8.00,40.00){\makebox(0,0)[rc]{$Y_0,\Theta_0$}}
\put(54.00,16.00){\makebox(0,0)[ct]{$S_c^{(0)}$}}
\put(43.00,39.00){\makebox(0,0)[cb]{$S_c$}}
\end{picture}
\caption{
   The choice of the local coordinates at the ends of correlated
   parts of  of both $8$ and $0$ like trajectories.
}
\label{fig:ythetas}
\end{figure}
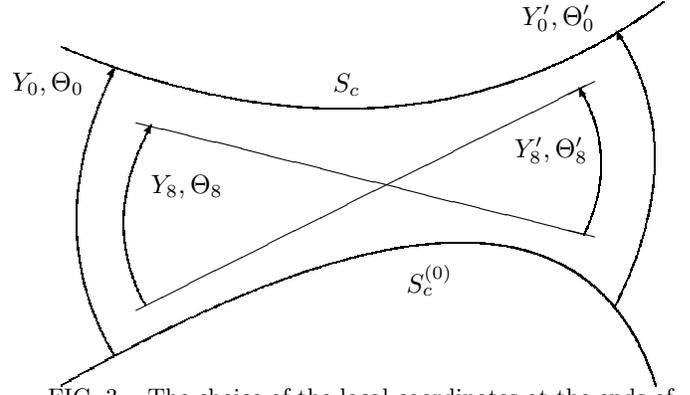

\section{ Some properties of 8-0 pairs. }
\label{sec:pspff}

The previous section shows how to construct 0-like trajectory for any 8-like
trajectory. The construction is valid if the expansions
Eqs.~(\ref{eq:con.1})
-- (\ref{eq:con.3}) are valid for $y$s given by Eq.~(\ref{eq:con.4}). This
is
always true for sufficiently long trajectories.  In this section we want to
setup relations between the period of trajectories and $\Delta S$.

The self-touching trajectory must have ``correlated''
part\cite{Aleiner-Larkin-96}. Indeed in order to be able to touch itself
trajectory must approach itself. In chaos convergence of trajectories (as
well as divergence) must be gradual with a rate known as Lyapunov exponent.
The piece where trajectory approaches itself and then leaves itself is
called
``correlated'' part of the loop.

The calculation of Sec.~\ref{sec:con} can be repeated in any crossection of
8-0 pair, where trajectory is sufficiently close to itself. The general
expression for action difference valid for any crossection is
\begin{equation}
    \Delta S = {1\over 2}
    ( pY_0\Theta_8 - pY_8\Theta_0) \;.
\label{eq:pspff.1}
\end{equation}
Here we changed notations of Sec.~\ref{sec:con}, $Y\rightarrow Y_0$,
$\Theta \rightarrow \Theta_8$, and added $Y_8$ - the distance between
opposite pieces of 8-like trajectory (it was zero in Sec.~\ref{sec:con}) and
$\Theta_0$ - the angle between pieces of 0-like orbit, see
Fig.~\ref{fig:ythetas}.  Let us now make two crossections of 8-0 pair near
boundaries between correlated and un-correlated parts. Expanding action of
this ``common'' piece of all trajectories,
\begin{equation}
   S_{c} = S_{c}^{(0)}
   + {1\over 2} m^c_{11} Y_0^2
   - m_{12}^{c} Y_0 Y_0'
   + {1\over 2} m^c_{22} Y_0'^2 \;,
\label{eq:pspff.2a}
\end{equation}
we can express action difference in terms of $Y$s :
\begin{equation}
   \Delta S = - {pm_{12}^c \over 2}
   ( Y_0 Y_8' - Y_0' Y_8 )
   \approx pm^c_{12} Y_0 Y_8
\label{eq:pspff.2}
\end{equation}
where  we put $ Y_8' \approx - Y_0 $. This is justified if the
``uncorrelated'' parts of trajectories are long enough and therefore
``soft''
enough (much more unstable then the correlated part).

The advantage of approximation Eq.~(\ref{eq:pspff.2})  is that the $\Delta
S$
is expressed in terms of one $0$-like part of the trajectory. Finally we
replace $m^c_{12}$ by the stability amplitude and arrive at
\begin{equation}
   \Delta S \approx  p
   { Y_0 \Theta_0' - Y_0' \Theta_0
    \over
    4\cosh^2(\lambda /2)
   }  \;,
\label{eq:pspff.3}
\end{equation}
where $\exp{\pm\lambda }$ are eigenvalues of the stability matrix of the
correlated part of the trajectory $2\cosh(\lambda) = (m_{11}^c +
m_{22}^c) / m_{12}^c$.  Here $Y$s and $\Theta$s belong to 0-like trajectory.
>From any of Eqs.~(\ref{eq:pspff.2}) or (\ref{eq:pspff.3}) we can derive
estimate $\Delta S <  \text{const  } e^{ - \tilde \lambda t } $,
where $\tilde \lambda$ is the stability exponent per unit time (Lyapunov
exponent) and $t$ is time the particle spend on the correlated part of the
trajectory. For ergodic systems there are infinitely many self-touching
trajectories with arbitrary long ``correlated'' part and therefore with
exponentially small $\Delta S$.


\section{ Operator form of the interference. }
\label{sec:opr}

The title of the work is divergence and interference in quantum chaos. The
term divergence is clear: take two particles very close initially in the
phase
space, let them propagate and they will branch off exponentially fast. Let
us
describe trajectories of the system by $\gamma(t) = F_t(\gamma(0))$, where
$\gamma$ is a point on the energy shell of the phase space.  The evolution
of
the two-particle distribution function is given by $f(\gamma_1,\gamma_2;t) =
\int d\gamma_3 d\gamma_4 {\cal A}(\{\gamma_j\};t) f(\gamma_3,\gamma_4;t) $
where
\begin{eqnarray}
   {\cal A}(\{\gamma_j\};t)  &=&
   \delta( \gamma_3 - F_{-t}(\gamma_1) )\;
   \delta( \gamma_4 - F_{-t}(\gamma_2) )
\label{eq:new.1}
\\
   &=&
   \delta( \gamma'_\parallel - \gamma_\parallel )\;
   \delta( \gamma'_\perp - \hat {\cal M} \gamma_\perp )\;
   \delta( \Gamma' - F_{-t}(\Gamma) )\;.
\nonumber
\\
\label{eq:new.2}
\end{eqnarray}
Here $\Gamma'=(\gamma_3+\gamma_4)/2$, $\Gamma =(\gamma_1+\gamma_2)/2$ are
centers of mass, $\gamma = \gamma_1- \gamma_2$, $\gamma' = \gamma_3 -
\gamma_4$ describe relative motion of two particles. Relative coordinates
are
separated to components which are either parallel or perpendicular to the
center-of-mass trajectory. We choose $ \gamma_{\perp} = (Y_0,\Theta_0)$
and $ \gamma'_{\perp}  = (Y_0',\Theta_0') $, see Fig.~\ref{fig:ythetas}. The
monodromy matrix $\hat {\cal M}$ can be expressed in terms of action
derivatives:
\begin{equation}
      \hat {\cal M}  =  {1\over m_{12}^c}
      \left( \begin{array}{cc}
          m_{11}^c  & 1 \\
          m_{11}^c m_{22}^c- (m_{12}^c)^2 & m_{22}^c
      \end{array}\right)
\label{eq:new.2a}
\end{equation}
Basically, the operator Eq.~(\ref{eq:new.2}) is eight-leg quantum
propagator.
The eight coordinates were  broken to four pairs and the Fourier transform
with respect of their difference in each pair leads to the density
propagator.

Somewhat different pairing of arguments of eight-leg propagator gives
interference operator. It is called the Hikami box in theory of disordered
metals\cite{Hikami-81}. We derived it for chaotic system by making use of
the
method of Aleiner and Larkin\cite{Aleiner-Larkin-96}. We just replaced 
random potential by exact scattering amplitudes. The interference operator
is 
similar to divergence operator:  it propagates center of mass along true 
trajectory but the relative motion is constrained in a different way:  
\begin{equation}
    {\cal H}(\{\gamma_j\};t)   =
    - e^{i\Delta S/\hbar}
    (m_{12}^c)^2
    \left[
      { {\cal A}(\{\gamma_j\};t) \over (m_{12}^c)^2}
    \right]_{ m_{12}^c  \rightarrow 0}
    \;.
\label{eq:new.5}
\end{equation}
This is main result of the present work.

The connection between interference and divergence becomes obvious. The
operator Eq.~(\ref{eq:new.5}) is important for trajectories with $\Delta S
\lesssim \hbar$. The crucial point is that $\Delta S$ is inversely
proportional to stability of the trajectory. Therefore for {\em any energy
of
the particle } there are sufficiently long and unstable trajectories
contributing to the interference kernel Eq.~(\ref{eq:new.5}).


\section{ Summary }
\label{sec:smm}

In summary we have given definition of interference in quantum chaos. It can
be seen as presence of nearly same length trajectories or as decrease of
transmission of some path and therefore increase of the return probability.
In this way we see that the trace formula do describes weak localization,
contrary to the statement of Ref.\onlinecite{Whitney-Lerner-Smith}.

The progress of non-perturbative field theoretical methods in quantum chaos
left the problem of interference open\cite{AAA-dec95,MK-jul95}. Indeed the
interference term in these theories has exactly the same structure as in
effective Lagrangian of disordered systems\cite{Efetov-jan83}. In latter
case interference between particles is point-like; it appears as a result of
single impurity scattering. In principle, one can add weak random potential
to chaotic system, derive effective Lagrangian\cite{Aleiner-Larkin-97} and
then put relaxation time to infinity.  Interference terms slowly disappear
together with random potential.

The form of the interference kernel is primarily important for building of
the
field theory correctly describing ``interaction'' of diffusion modes (
called
in quantum chaos Liuvillian modes). At present time it is not clear how to
derive theory with interference kernel Eq.~(\ref{eq:new.5}) in the effective
action.

The developed here theory can be generalized to any dimension, however the
counting of phase lost due to beam rotation becomes more complicated.
In addition, the  theory is not sensitive to any discrete symmetries and can
be applied to modular groups\cite{Bogomolny-Leyvraz-Schmit}. When
the manuscript was in prpeapration author received mail about
similar calculation of the action difference of 8-0 pairs undertaken
recently by other people.\cite{Sieber-Richter}

Comments and remarks of D. Cohen are greatly acknowledged.


\end{document}